\documentclass{svproc}
\usepackage{amssymb,amsmath,amsfonts,mathtools}
\usepackage{cite} %makes the citations in the form [1-5]
\usepackage{graphicx}
\usepackage{epsfig}
\usepackage{tikz}
\usetikzlibrary{calc}

\newcommand{\sub}[1]{_{\stackrel{}{#1}}}

\newcommand{\ben}{\begin{equation}}
\newcommand{\een}{\end{equation}}
\newcommand{\bea}{\begin{eqnarray}}
\newcommand{\eea}{\end{eqnarray}}

\newcommand{\QQ}{\qquad\qquad\qquad\qquad}

\newcommand{{\ga}}{{\gamma}}

\newcommand{\cB}{{\cal B}}

\newcommand{\ch}{{\mathrm{ch}}}

\newcommand{\mult}{{\mathrm{mult}}}

\def\Z{{\Bbb Z}} \def\N{{\Bbb N}}
\def\R{{\Bbb R}}

\usepackage{url}

\begin{document}
\mainmatter              % start of a contribution
\title{Demazure Formulas for Weight Polytopes}
\titlerunning{Demazure Formulas for Weight Polytopes}  % abbreviated title (for running head)
\author{Mark A. Walton}
\authorrunning{M.A. Walton} % abbreviated author list (for running head)
%Craig Chambers, Kim B. Bruce, and Elisa Bertino}
%
\institute{Department of Physics \& Astronomy, University of Lethbridge\\  
Lethbridge, Alberta, Canada T1K 3M4\\ 
\email{walton@uleth.ca}  }
%\\ WWW home page: \texttt{http://users/\homedir iekeland/web/welcome.html}
%\and
%Universit\'{e} de Paris-Sud,
%Laboratoire d'Analyse Num\'{e}rique, B\^{a}timent 425,\\
%F-91405 Orsay Cedex, France}

\maketitle              % typeset the title of the contribution

\begin{abstract}
The characters of simple Lie algebras are naturally decomposed into lattice polytope sums. The Brion formula for those polytope sums is remarkably similar to the Weyl character formula. Here we start to investigate if other character formulas have analogs for lattice polytope sums, by focusing on the Demazure character formulas.  Using Demazure operators, we write expressions for the lattice sums of the weight polytopes of rank-2 simple Lie algebras, and the rank-3 algebra $A_3$. 
% We would like to encourage you to list your keywords within
% the abstract section using the \keywords{...} command.
\end{abstract}
\section{Introduction} 

The Brion formula \cite{{Brion88}, {Brion92}} for lattice polytope sums is remarkably similar to the Weyl formula for characters of simple Lie algebras \cite{DhillonKhare18, Walton04, Postnikov09}.  As a consequence, the expansion of Weyl characters in terms of lattice polytope sums is natural and useful \cite{Walton04, Postnikov09, Schutzer12, Walton13, DhillonKhare18, Rasmussen18}.  This {\it polytope expansion} of Lie characters is highly reminiscent of the early work of Antoine and Speiser \cite{AntoineSpeiser64} and the recursive formulas found by Kass \cite{Kass91}. 

Here we explore further the relation between Lie characters and lattice polytope sums. Other formulas exist for the characters - might corresponding formulas describe lattice polytope sums?\footnote{This question was already asked in \cite{Walton04}.}  

We focus on the Demazure character formulas \cite{Demazure74, Andersen85, Joseph85}, and use the Demazure operators involved to write expressions for the lattice polytope sums.  So far, we have obtained results for all rank-2 simple Lie algebras, and for the rank-3 algebra $A_3$.  We hope these first formulas will help lead to general Demazure formulas for the lattice polytope sums relevant to Lie characters, and perhaps others. 

How might such a formula be useful? Starting with the Demazure character formulas, Littelmann was able to derive a generalization for all simple Lie algebras of the famous  Littlewood-Richardson rule for $A_r$ tensor-product decompositions  \cite{Littelmann90}.  In a similar way, formulas of the Demazure type might  lead to efficient, general computational methods for lattice polytopes. 

Physical applications should also be possible. An attempt to apply the Demazure character formula to Wess-Zumino-Witten conformal field theories was made in \cite{Walton98}, and it has already been used in the study of solvable lattice models \cite{Kuniba98}. 

In the following section, we review the initial motivation for the present work, the similarity between the Weyl character formula and  the Brion lattice-polytope sum formula, and the polytope expansion that exploits it. Section 3 is a quick review of the Demazure character formulas.  Our new results are presented in Section 4: expressions involving Demazure operators for lattice-polytope sums for rank-2 simple Lie algebras, and $A_3$.  The final section is a short conclusion. 

\vskip20pt
\section{Polytope expansion of Lie characters}
 
Let $X_r$ denote a simple Lie algebra of rank $r$, so that $X=A,B,C,D,E,F,$ or $G$.  The sets of fundamental weights and simple roots will be written as $F:=\{\, \Lambda^i\,\vert\, i\in \{1,2,\ldots, r\}\, \}$ and $S:=\{\, \alpha_i\,\vert\, i\in \{1,2,\ldots, r\}\, \}$, respectively. The weight and root lattices are $P:=\Z\, F$ and $Q:= \Z\, S$, respectively.  The set of integrable weights is $P_+ := \N_0\, F$, and we write $R$ ($R_+$) for the set of (positive) roots of $X_r$. 

\vskip10pt
\subsection{Weyl character formula} 

Consider an irreducible representation $L(\lambda)$ of $X_r$ of highest weight $\lambda\in P_+$. 
The formal character of $L(\lambda)$ is defined to be 
\begin{align}
\ch_\lambda\ =\ \sum_{\mu\in P(\lambda)}\, \mult_\lambda(\mu)\, e^{\,\mu}\, .
\label{char}\end{align}
Here $P(\lambda)$ is the set of weights of the representation $L(\lambda)$, and $\mult_\lambda(\mu)$ is the multiplicity of weight $\mu$ in $L(\lambda)$.  

The formal exponentials of weights obey $e^\mu\, e^\nu\, =\, e^{\mu + \nu}$.  If we write 
\begin{align}
e^\mu(\sigma)\ =:\ e^{\langle \mu, \sigma\rangle}\  ,
\label{FormalExp}
\end{align}
where $\langle \mu, \sigma\rangle$ is the inner product of  weights $\mu$ and $\sigma$, the formal exponential $e^\mu$ simply stands for $e^{\langle \mu, \sigma\rangle}$ before a choice of weight $\sigma$ is made.  A choice of $\sigma$ fixes a conjugacy class of elements in the Lie group $\exp(X_r)$.  The formal character then becomes a true character $\ch_\lambda(\sigma)$, the trace,  in the representation of highest weight $\lambda$, of elements of $\exp(X_r)$ in the conjugacy class labelled by  $\sigma$. 
 
The celebrated Weyl character formula is 
\begin{align}
\ch_\lambda\ =\ \frac{\sum_{w\in W} (\det w)\, e^{w.\lambda}}{\prod_{\alpha\in R_+}\, (1-e^{-\alpha})}\ .
\label{WCF}\end{align} 
Here $W$ is the Weyl group of the simple Lie algebra $X_r$, and $w.\lambda = w(\lambda+\rho)-\rho$ denotes the shifted action of Weyl-group element $w\in W$ on the weight $\lambda$,  with $\rho=\sum_{i=1}^r{\Lambda^i}$.  

The Weyl invariance of the character can be made manifest: 
 \begin{equation}
{\rm ch}_\lambda\ =\
\sum_{w\in W}\, e^{w\lambda}\, \prod_{\alpha\in  R_+}\,
(1-e^{-w\alpha})^{-1}\ . 
\label{WCFformal}\end{equation}
Here now 
\begin{equation}
(1-e^\beta)^{-1}\ = \left\{\begin{matrix} 1+e^\beta+e^{2\beta}+\ldots\ &,\ \beta\in R_-; \cr  -e^{-\beta}-e^{-2\beta}-\ldots &,\  \beta\in R_+. \end{matrix} \right. 
\label{series}\end{equation}
The rule-of-thumb is: expand in  powers of $e^\beta$, with $\beta$ a negative root.

\vskip10pt
\subsection{Brion lattice-polytope sum formula}

A polytope is the convex hull of finitely many points in $\R^d$. A polytope's vertices are such a set of points with minimum cardinality. A lattice polytope has all its vertices in an integral lattice in $\R^d$.  The 
(formal) lattice polytope sum is the sum of terms $e^{\,\phi}$ over the lattice points $\phi$ in the polytope. 

Brion \cite{Brion88, Brion92} found a general formula for these lattice-polytope sums.  Let the {\textit{weight polytope}} ${\rm Pt}_\lambda$ be the polytope with vertices given by the Weyl orbit $W\lambda$.  Consider the lattice-polytope sum 
\begin{equation}   
{\rm B}_\lambda\ :=\ \sum_{\mu\in (\lambda+Q)\cap{\rm Pt}_\lambda}\, e^{\, \mu}\ ,\label{Bdefinition}\end{equation}
where the relevant lattice is the $\lambda$-shifted root lattice $\lambda + Q$ of the algebra $X_r$.  Applied to a weight polytope, the Brion formula yields
\begin{equation}
{\rm B}_\lambda\ =\
\sum_{w\in W}\, e^{w\lambda}\, \prod_{{\alpha\in }S}\,
(1-e^{-w\alpha})^{-1}\ . 
\label{Brion}\end{equation}
Here $S$ denotes the set of simple roots of $X_r$.

\vskip10pt
\subsection{Polytope expansion}
 
The Brion formula (\ref{Brion}) is remarkably similar to the Weyl character formula, as written in (\ref{WCFformal}) \cite{Walton04, DhillonKhare18}.  It is therefore natural, and fruitful, to consider the polytope expansion of Lie characters \cite{Walton04, DhillonKhare18, Walton13}:  
\begin{equation}  {\rm ch}_\lambda\ =\ \sum_{\mu\leq\lambda}\, {\rm polyt}_\lambda(\mu)\,\, {\rm B}_\mu\ .\ 
\label{PolytExpn}\end{equation} 
The polytope multiplicities ${\rm polyt}_\lambda(\mu)$ are defined in analogy with weight multiplicities ${\rm mult}_\lambda(\mu)$. 

We do not consider the polytope epansion further in this note. Instead we focus on the striking relation described above between characters and polytope sums.

\vskip20pt
\section{Demazure character formulas}

Do other character formulas point to the existence of new formulas for the lattice sums of weight polytopes?  More general polytopes?

In particular, do the Demazure formulas for Lie characters indicate the existence of Demazure-type formulas for the lattice sums of weight polytopes? 

Let us first sketch the Demazure character formulas. The Weyl group $W$ is generated by the primitive reflections $r_i$  in weight space across the hyperplanes normal to the corresponding simple roots $\alpha_i$:  
\begin{align}
r_i\,\lambda\ =\ \lambda\ -\ (\lambda\cdot\alpha^\vee_i)\, \alpha_i\ ;
\label{PrRefl}
\end{align}
here $\alpha^\vee_i = 2\alpha_i/(\alpha_i\cdot \alpha_i)$ is the simple co-root. 

Define Demazure operators for every simple root $\alpha_i\in S$:
\begin{align}
D_{\alpha_i}\ =:\ D_i\  =\ \frac{1\ -\ e^{-\alpha_i}\,  r_i}{1-e^{-\alpha_i}}\ ,
\label{Diri}
\end{align}
where 
$r_i(e^\lambda)\ =\ e^{r_i \lambda}\ .$  Then 
\begin{align}
D_i\, e^{\,\lambda}\ =\ \left\{\begin{matrix}  e^{\, \lambda}+ e^{\,\lambda-\alpha_i} + e^{\,\lambda-2\alpha_i} + \ldots + e^{\, r_i\lambda}, &\quad \lambda\cdot\alpha_i^\vee \ge 0\ ;\cr  -e^{\, \lambda+\alpha_i} - e^{\,\lambda+2\alpha_i} - \ldots - e^{\, r_i(\lambda+\alpha_i)}, &\quad \lambda\cdot\alpha_i^\vee < 0\ .\end{matrix}\right.
\label{Di}
\end{align}

For every Weyl-group element $w\in W$ a Demazure operator $D_w$ can be defined.  First, identify 
$D_{r_i}:=D_i$, and then use any reduced decomposition of $w$, replacing the factors $r_j$ in the reduced decomposition with $D_j$.  The resulting operator $D_w$ must be independent of which reduced decomposition is used. As a result, the Demazure operators obey relations encoded in the Coxeter-Dynkin diagrams of $X_r$. 

For example, consider the longest element $w_L$ of the Weyl group of $A_2$: 
$w_L\ =\ r_1r_2r_1\ =\ r_2r_1r_2$. The associated Demazure operator $D_{w_L}$ can be written two ways, so that $D_1D_2D_1\ =\ D_2D_1D_2.$

If $w_L$ is the longest element of the Weyl group $W$, then the Demazure character formula is 
\begin{equation}  
\ch_\lambda\ =\ D_{w_L}\, e^{\, \lambda}\ .
\label{DemazureD}\end{equation}

Also, define $D_i\ =:\ 1\ +\ d_i$, and then $d_w$ for all $w\in W$ by reduced decompositions. Then
\begin{equation}\ch_\lambda\ =\ \sum_{w\in W}\, d_{w}\, e^{\, \lambda}\ . 
\label{Demazured}\end{equation}

Demazure operators can also be defined for every positive root $\beta\in R_+$: 
\begin{align}
D(\beta)\   :=\ \frac{1\ -\ e^{-\beta}\,  r(\beta)}{1-e^{-\beta}}\ ,
\label{Dbrb}
\end{align}
where 
\begin{align}
r(\beta)\lambda\ :=\ \lambda\, -\, (\lambda\cdot\beta^\vee)\beta\  {\rm\ \ and\ \ \ } r(\beta)(e^\lambda)\ =\ e^{r(\beta)\lambda}\ .
\label{rb}
\end{align}
Operators $d(\beta)\, =\, D(\beta)\, -\, 1$, 
\begin{align}
d(\beta)\   :=\ \frac{e^{-\beta}\,\left[ 1\, -\,   r(\beta)\right]}{1-e^{-\beta}}\ ,
\label{dbrb}
\end{align} 
are also defined for all positive roots $\beta\in R_+$.

\vskip20pt
\section{Lattice-polytope formulas of Demazure type}

In the hopes of helping lead to a more general result, we take a direct approach here, and write formulas for low-rank weight-polytope sums that involve the Demazure operators.  We report only preliminary new results, formulas for all rank-2 cases, and for one of rank 3, related to the Lie algebra $A_3$. 

But before treating ranks 2 and 3, let us first dispense with the unique rank-1 algebra, $A_1$.  In this case, the character and weight-polytope lattice sum are identical, \begin{align}
{\rm {ch}}_\lambda\ =\  B_\lambda\ =\ e^\lambda + e^{\lambda-\alpha_1} + e^{\lambda-2\alpha_1}+\ldots + e^{-\lambda}\ .
\label{AiBch}
\end{align}
The Demazure character formulas therefore apply to $B_\lambda$. 

The $A_1$ Weyl group $W$ has 2 elements, the identity and $r_1 = r(\alpha_1)$, where $\alpha_1$ is the simple root, and only positive root. The longest element of $W$ is therefore $w_L = r_1 = r(\alpha_1)$, with a unique reduced decomposition. Applying the Demazure formulas (\ref{DemazureD}, \ref{Demazured}), we find 
\begin{align}
B_{\lambda}\ =\ D_1\, e^\lambda\ =\ \big[ d(\alpha_1)\, +\, 1 \big]\, e^\lambda\ . 
\label{AiBDd}
\end{align}
The last expression will turn out to be the most relevant here - see (\ref{RankTwoB}, \ref{AThreeB}) below.

The 3 rank-2 algebras can be treated in a unified way. Put the $p:= \dim{R_+}$ positive roots of your rank-2 algebra in angular order; label them $\gamma_j$. So we get, for $A_2$, 
\begin{align}
\{\gamma_1, \gamma_2, \gamma_{p=3} \}\ =\ \{\alpha_1, \alpha_1+\alpha_2, \alpha_2 \}\ ;
\end{align}
for $B_2$ ($\cong C_2$), 
\begin{align}
\{\gamma_1, \ldots , \gamma_{p=4} \}\ =\ \{\alpha_1, \alpha_1+\alpha_2, \alpha_1+2\alpha_2, \alpha_2 \}\ ;
\end{align}
and for $G_2$, 
\begin{align}
\{\gamma_1, \ldots , \gamma_{p=6} \}\ =\ \{\alpha_1, \alpha_1+\alpha_2, 2\alpha_1+3\alpha_2, \alpha_1+2\alpha_2, \alpha_1+3\alpha_2, \alpha_2 \}\ .
\end{align}
In a generic weight diagram, these positive roots are parallel to half of the boundaries, in angular order.  They specify a path from the highest weight to the lowest weight along the polytope edges labelled by $\gamma_1$ through $\gamma_p$, in that order. Correspondingly, the longest element $w_L$ of the Weyl group can be written as a product of the reflections defined in (\ref{rb}): 
\begin{align}
w_L\ =\ r(\gamma_p) r(\gamma_{p-1})\, \cdots\, r(\gamma_2) r(\gamma_1)\ . 
\label{wLgammas}
\end{align}

The weights of the first $p-1$ boundaries can be generated, and then the polytope can be filled in by $\gamma_p$-strings of weights.  If $B_\lambda\ =\ \cB\,( e^\lambda)$, then
\begin{align}
\cB\, =\, \big[d({\gamma\sub{p}}) + 1\big]\, \Big[\, d(\gamma\sub{{p-1}}) r(\gamma\sub{{p-2}})\cdots r(\gamma\sub{{1}})&\, +\, \QQ\quad\cr d(\gamma\sub{{p-2}}) r(\gamma\sub{{p-3}})\cdots r(\gamma\sub{{1}})\, +\, \ldots\, +\,& d(\gamma\sub{{2}}) r(\gamma\sub{{1}})\, +\,d(\gamma\sub{{1}})\, +\, 1\, \Big] . 
\label{RankTwoB}
\end{align}
Here we use the Demazure operators defined in (\ref{dbrb}) above. Eqn.~(\ref{RankTwoB}) is one formula of Demazure type that applies to all rank-2 cases. 

Now consider a rank-3 example, the Lie algebra $A_3$. The weight polytope for a highest weight with all Dynkin labels non-zero, and unequal, is illustrated in Figure 1. Notice that the facets are the weight polytopes for the rank-2 algebras whose Coxeter-Dynkin diagrams are obtained from that of $A_3$ by deleting a single node. Hence, the facets are hexagonal $A_2$ weight polytopes of 2 types, and rectangular $A_1\oplus A_1$ polytopes. 

\begin{figure}[ht]%[scale=1.]
\hskip+1.5cm
\includegraphics[scale=0.7,clip]{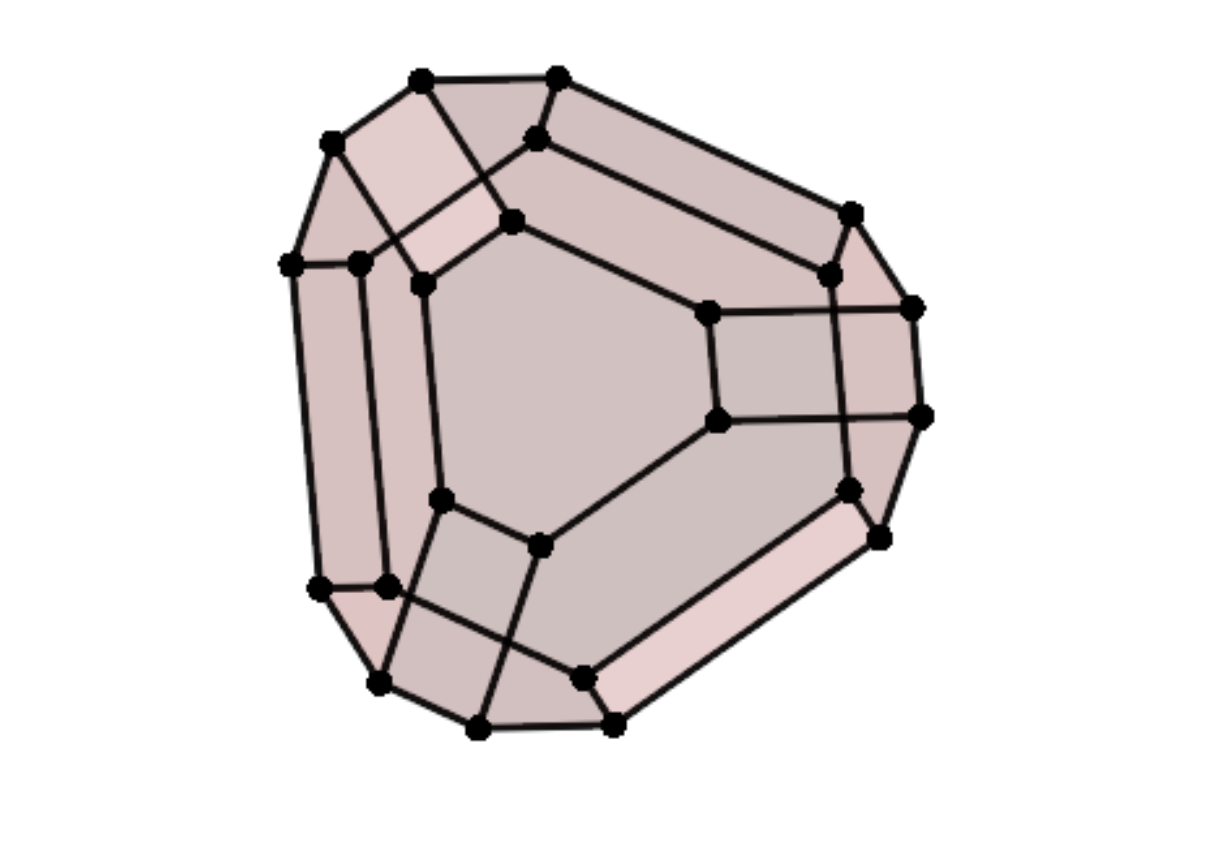}
\vskip-0.5cm
\caption{Weight polytope for a regular highest weight of the Lie algebra $A_3$.}
\label{figure:RootsG2}
\end{figure}

The longest element of the $A_3$ Weyl group can be written as in (\ref{wLgammas}), with $p=6$ and 
\begin{align}
\{\gamma_1,\ldots,\gamma_6\}\ =\ \{ \alpha_1, \alpha_{12}, \alpha_{123}, \alpha_2, \alpha_{23}, \alpha_3 \}\ ,
\end{align}
where $\alpha_{12}:=\alpha_1+\alpha_2$, etc.  The expression corresponds to a path along the edges of the polytope from the highest to the lowest weight. 

It is not difficult to see that the following expression 
\begin{align}
\cB\, =\, \big[d(\alpha_{3})\, &+\, 1\big]\,  \big[d(\alpha_{23})r(\alpha_2)\, +\, d(\alpha_{2})\, +\, 1\big]\,\, \cr \big[&d(\alpha_{123})r(\alpha_{12})r(\alpha_1)\, +\, d(\alpha_{12})r(\alpha_1)\, +\, d(\alpha_1)\, +\, 1 \big] 
\label{AThreeB}
\end{align}
generates the $A_3$ weight-polytope lattice sum. Notice the order of factors follows the order of factors in the expression (\ref{wLgammas}), as in the rank-2 result (\ref{RankTwoB}).

\vskip20pt
\section{Conclusion} 

We have begun a search for Demazure-type formulas for the exponential sums of weight polytopes of simple Lie algebras.  Our results are preliminary, mostly summarized in (\ref{RankTwoB}) and (\ref{AThreeB}), expressions valid for all rank-2 algebras ($A_2, B_2\cong C_2, G_2$) and $A_3$, respectively.  

Clearly, Demazure-type formulas can be written.  However, our expressions are not unique - we have obtained others. What is needed is a universal formula, one that applies to all simple Lie algebras, as the Weyl (\ref{WCF}, \ref{WCFformal}) and Demazure (\ref{DemazureD}, \ref{Demazured}) character formulas do, as well as the Brion formula (\ref{Brion}) for a weight polytope. Similarities in the formulas given here may indicate that we are on the right track. 

\vskip0.75cm\section*{{Acknowledgements}}
\vskip-0.2cm
I thank J\o rgen Rasmussen for collaboration and Chad Povey for 3D-printing rank-3 weight polytopes. This research was supported by a Discovery Grant from the Natural Sciences and Engineering Research Council of Canada (NSERC). 

\vskip30pt
\bibliographystyle{amsplain}

\begin{thebibliography}{10} 


\bibitem{Andersen85} H.H. Andersen, \textit{Schubert varieties and Demazure's character formula}, Invent. math. 79 (1985) 611-618. 
 
\bibitem {AntoineSpeiser64} 
J.-P. Antoine, D. Speiser,  
\textit{Characters of irreducible representations of the simple groups. 
I. General theory.}  
J. Mathematical Phys. {\bf 5} (1964) 1226--1234; \hfill\break 
\textit{Characters of irreducible representations of the simple groups. II. 
Application to classical groups},  
J. Mathematical Phys. {\bf 5} (1964) 1560--1572.


\bibitem{Brion88} M. Brion, \textit{Points entiers dans les poly\`edres convexes}, Ann. scient. \'Ec. Norm. Sup., 4e s\'erie, t. 21 (1988) 653-663.

\bibitem {Brion92} M. Brion, 
\textit{Poly\`edres et r\'eseaux}, 
Enseign. Math. (2) \textbf{38} (1992), no. 1-2, 71--88. 

\bibitem{Demazure74} M. Demazure, \textit{D\'esingularisation des vari\'et\'es de Schubert g\'en\'eralis\'ees}, Ann. scient. \'Ec. Norm. sup., t. 6,
Sect. 2 (1974) 53-88; \textit{Une nouvelle formule des caract\`eres}, Bull. Sci. Math. (2) 98 (1974), no. 3, 163-172.

\bibitem{DhillonKhare18} G. Dhillon, A. Khare, 
\textit{Characters of highest weight modules and integrability}, 
arXiv:1606.09640 (2016);  
\textit{The Weyl-Kac weight formula}, 
S\'eminaire Lotharingien de Combinatoire \textbf{78B} (2017), Proceedings of the 29th Conference on Formal Power
Series and Algebraic Combinatorics (London), Article \#{77},  
 arXiv:1802.06974 (2018).
 
\bibitem{Joseph85} A. Joseph, \textit{On the Demazure character formula}, Ann. sclent. \'Ec. Norm. Sup., 4${}^{\rm e}$ s\'erie, t. 18 (1985) 389 -419.
 
\bibitem{Kass91} S. Kass, 
\textit{A recursive formula for characters of simple Lie algebras}, 
J. Alg. 137 (1991) 126. 

\bibitem{Kuniba98} A. Kuniba, K. Misra, M. Okado, T. Takagi, J. Uchiyama, \textit{Characters of Demazure modules and solvable lattice models}, Nucl. Phys. B 510 (1998) 555-576. 

\bibitem{Littelmann90} P. Littelmann, 
\textit{A generalization of the Littlewood-Richardson rule},  J. Alg. 130 (1990) 328–368. 

\bibitem{Postnikov09} A. Postnikov, 
\textit{Permutohedra, associahedra, and beyond}, Int. Math. Res. Not. IMRN 6 (2009)  1026-1106.


\bibitem{Rasmussen18} J. Rasmussen,  
 \textit{Layer structure of irreducible Lie algebra modules}, preprint 
 arXiv:1803.06592 (2018).

\bibitem{Schutzer12} W. Schutzer, 
\textit{A new character formula for Lie algebras and Lie groups}, J. Lie
Theory, 22.3 (2012) 817-838. 

\bibitem{Walton98} M.A. Walton,
\textit{Demazure Characters and WZW Fusion Rules}, J. Math. Phys. 39 (1998) 665-681. 

\bibitem{Walton04} M.A. Walton, \textit{Polytope sums and Lie characters}, Symmetry in physics, volume 
34 of CRM Proc. Lecture Notes (AMS, 2004) 203-214, Proceedings of a CRM  Workshop held in Memory of Robert T. Sharp, 12-14 September 2002. 

\bibitem{Walton13} M.A. Walton, 
\textit{Polytope expansion of Lie characters and applications}, J. Math. Phys. 54 (2013) 121701.  








\end{thebibliography}

\end{document}